\documentclass{aa}
\usepackage{graphicx}
\usepackage{epsfig}
\def\ltsima{$\; \buildrel < \over \sim \;$}
\def\gtsima{$\; \buildrel > \over \sim \;$}
\def\simlt{\lower.5ex\hbox{\ltsima}}
\def\simgt{\lower.5ex\hbox{\gtsima}}

\begin{document}

   \title{X--ray polarization properties of the accretion column in magnetic CVs}

   \author{Giorgio Matt
          }

   \offprints{G. Matt}

   \institute{Dipartimento di Fisica, Universit\`a degli Studi Roma Tre,
via della Vasca Navale 84, I-00146 Roma, Italy\\
             }

   \date{Received ; accepted }

   \abstract{
The accretion column in magnetic Cataclysmic Variables may have a not negligible Thomson
optical depth. A fraction of the thermal radiation from the post--shock region may therefore
be scattered -- and then polarized -- before escaping the column. Moreover, part of the thermal
radiation is reflected -- and again polarized -- by the White Dwarf surface. We show that 
X--ray polarimetry can provide valuable, and probably unique informations on
the geometry and physical parameters of the accretion column by calculating,
by means of Monte Carlo simulations, the expected polarization properties of magnetic CVs
as a function of the geometrical parameters (assuming a cylindrical geometry)
and the Thomson optical depth of the column. 
We find that degrees of polarization as high as about 4\% can be present, and  apply our
calculations to the archetypal magnetic CV, AM~Herculis.
   }

   \maketitle
%

\section{Introduction}

The magnetic field in some subclasses of Cataclysmic Variables (notably in Polars and 
in at least a fraction of intermediate Polars; see Warner 1995 for a complete overview
on CVs) is strong enough to channel the accreting matter along the field lines. For
a dipolar field, this means that the accretion occurs via accreting columns on one or both
the magnetic poles; the disalignement between the spin and magnetic axes results in 
pulsed emission. 

Hard X--rays are then produced by optically thin thermal line and continuum emission
 in the so--called post--shock region, where temperatures can reach values as large as
several tens of keV (e.g. Frank et al. 1992; Cropper et al. 1999). Thermal emission is expected
to be unpolarized (see next section); however, the Thomson optical depth in the accretion
column, while probably less than unity, may be not negligible. Indeed, Hellier et al. (1998)
found significant broadening in the iron K$\alpha$ line of several magnetic CVs, which they
interpreted as due to Compton broadening. Thomson scattered radiation is polarized, provided
that the geometry is not spherical. It is therefore to be expected 
that the hard X--ray emission in magnetic CVs is polarized, with the net polarization degree
increasing with the Thomson optical depth of the accretion column. 

In this paper we calculate, by means of Monte Carlo simulations, the polarization properties
of the accretion column in magnetic CVs. In Sec. 2 the numerical
code is described, while the results
(including an application to the archetypal Polar, AM~Herculis) are presented in Sec.~3. 
Results are then summarized, and observational perspectives discussed, in Sec.~4.

\begin{figure*}[t]
\hbox{
\includegraphics*[width=8.0cm,height=8.0cm]{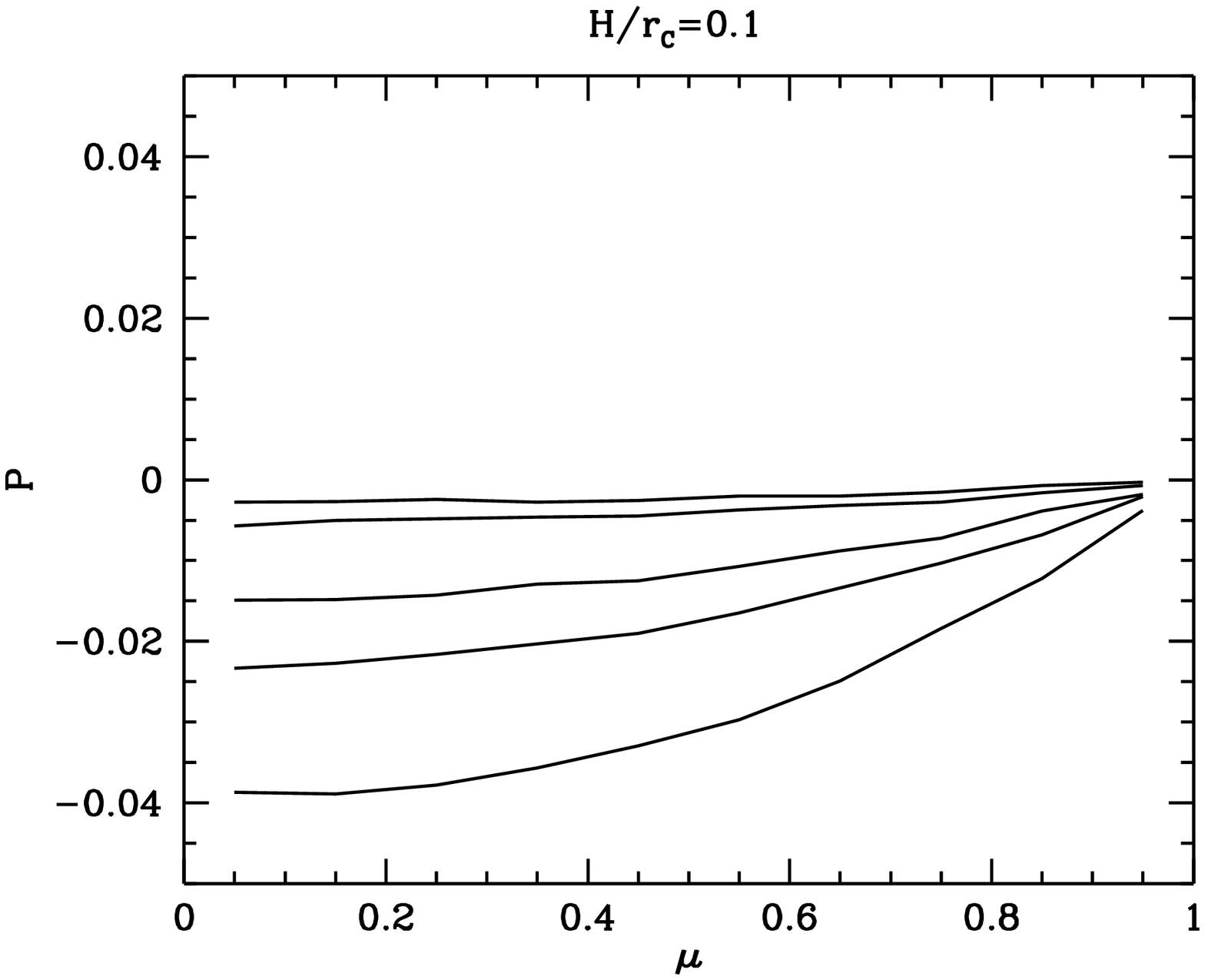}
\hspace{0.5cm}
\includegraphics*[width=8.0cm,height=8.0cm]{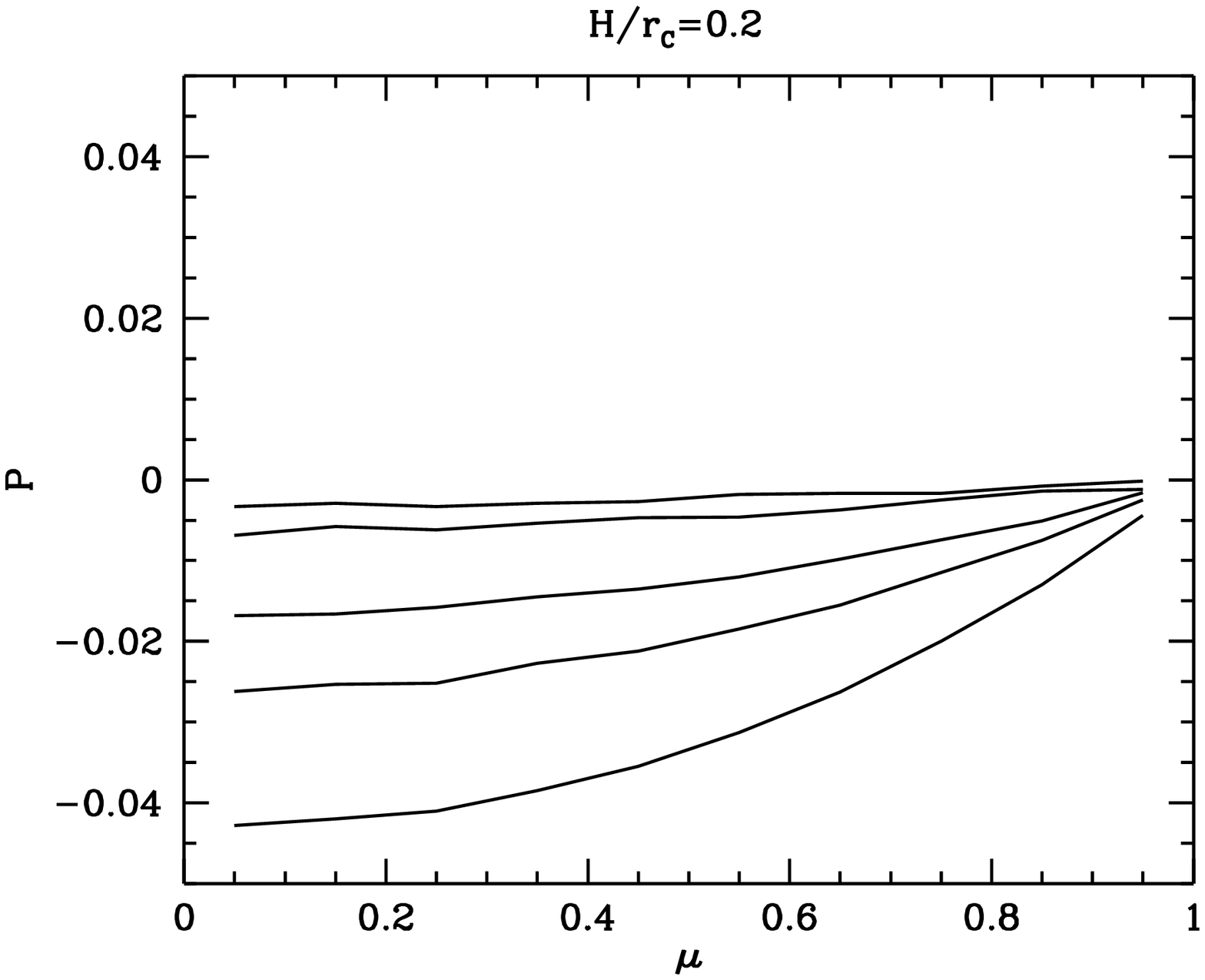}
}
\hbox{
\includegraphics*[width=8.0cm,height=8.0cm]{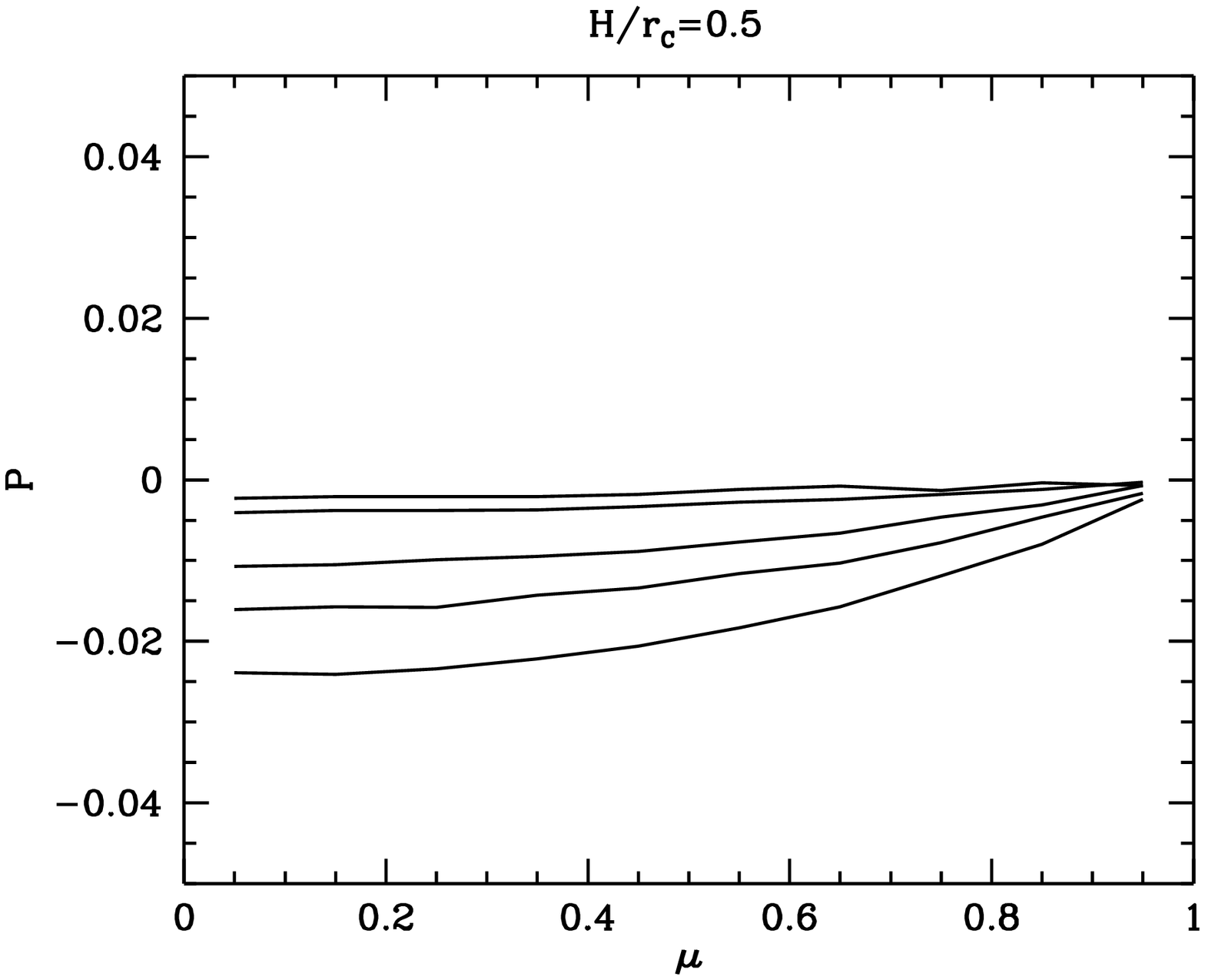}
\hspace{0.5cm}
\includegraphics*[width=8.0cm,height=8.0cm]{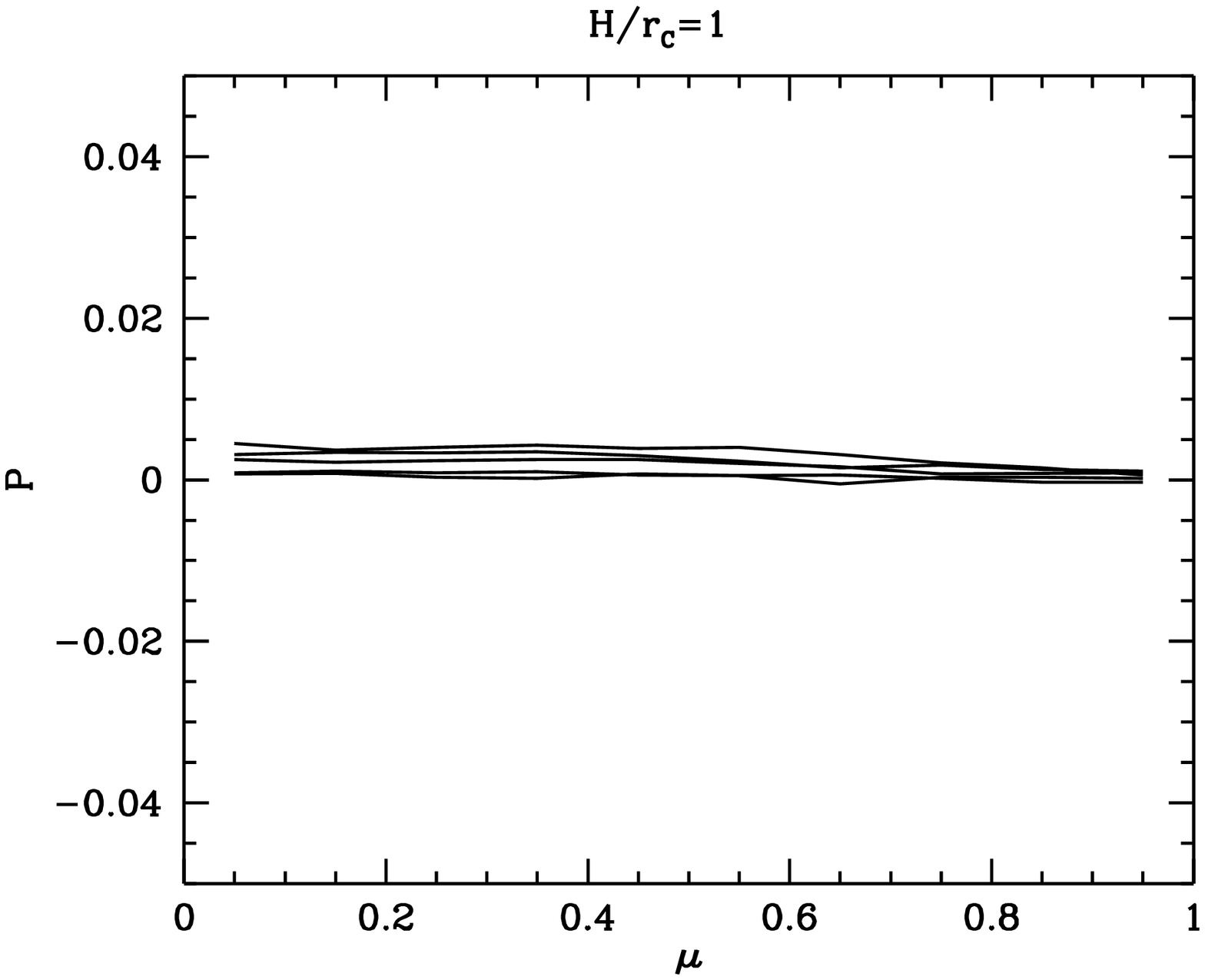}
}
\caption{The degree of polarization, P,  as a function of $\mu=cos\theta$ for four values
of $H/r_C$. In each panel, five values of $\tau_T$ are shown: 0.05, 0.1, 0.3, 0.5 and 1
(in increasing order of the absolute value of $P$). 
}
\label{polittico_1}
\end{figure*}

\begin{figure*}[t]
\hbox{
\includegraphics*[width=8.0cm,height=8.0cm]{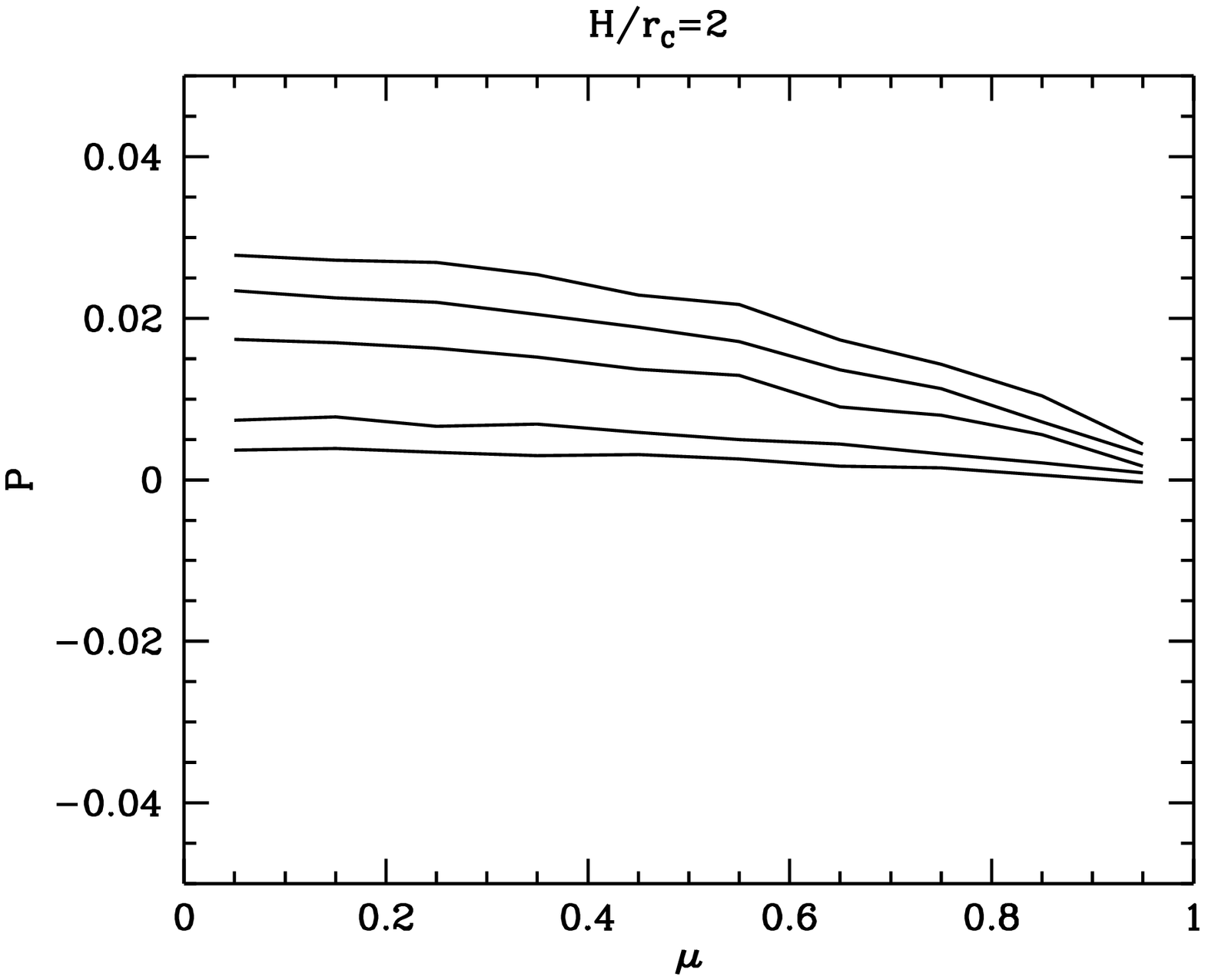}
\hspace{0.5cm}
\includegraphics*[width=8.0cm,height=8.0cm]{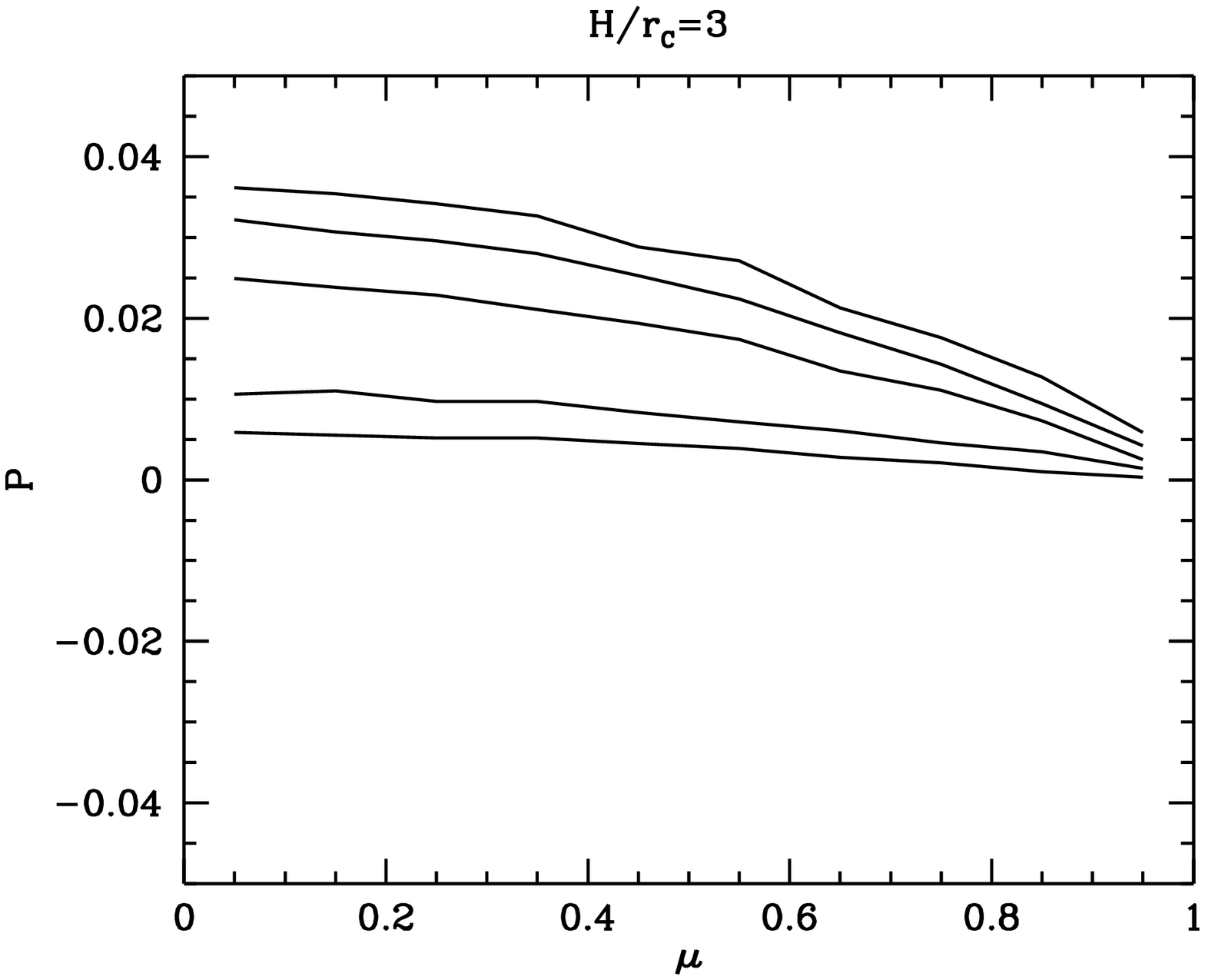}
}
\hbox{
\includegraphics*[width=8.0cm,height=8.0cm]{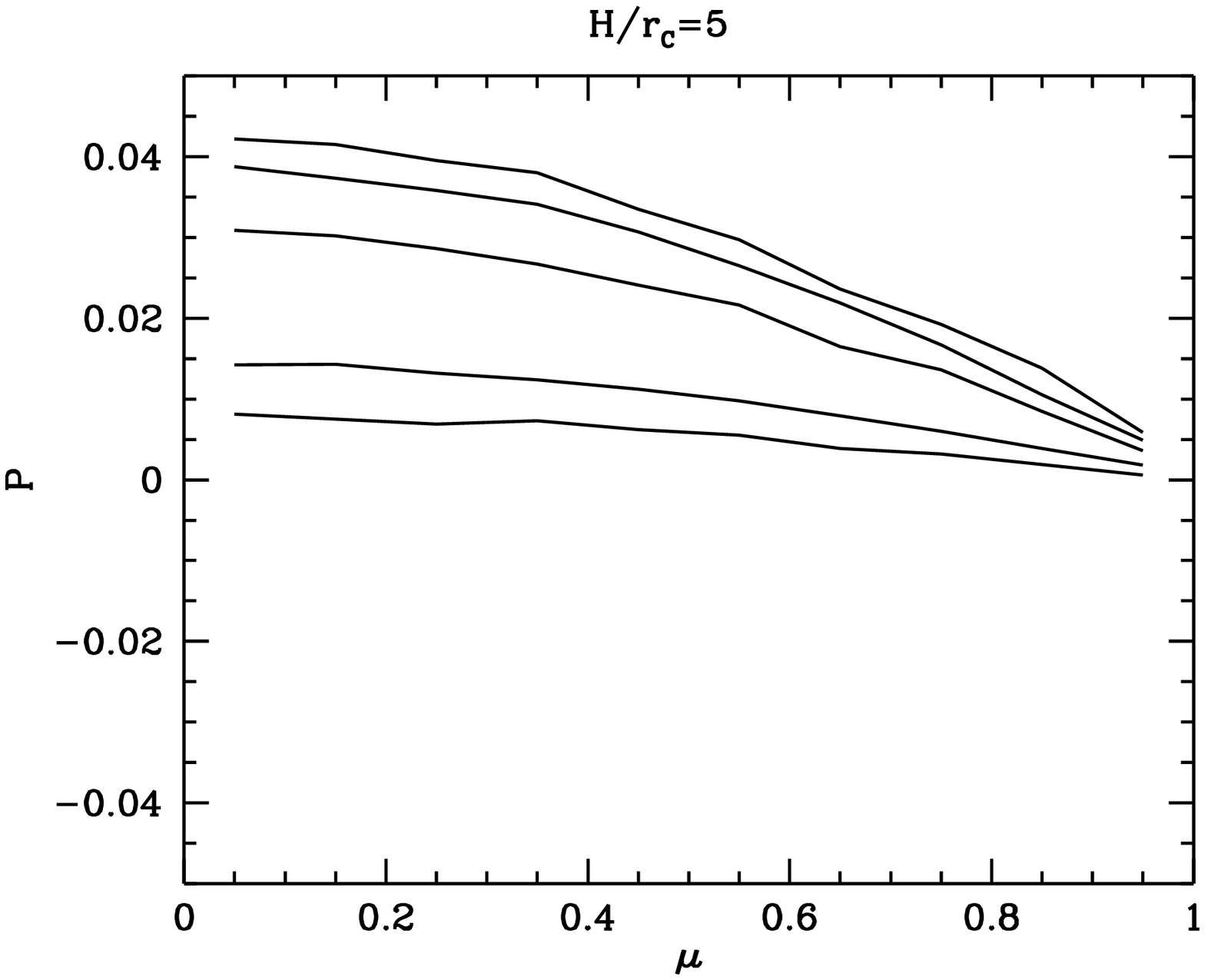}
\hspace{0.5cm}
\includegraphics*[width=8.0cm,height=8.0cm]{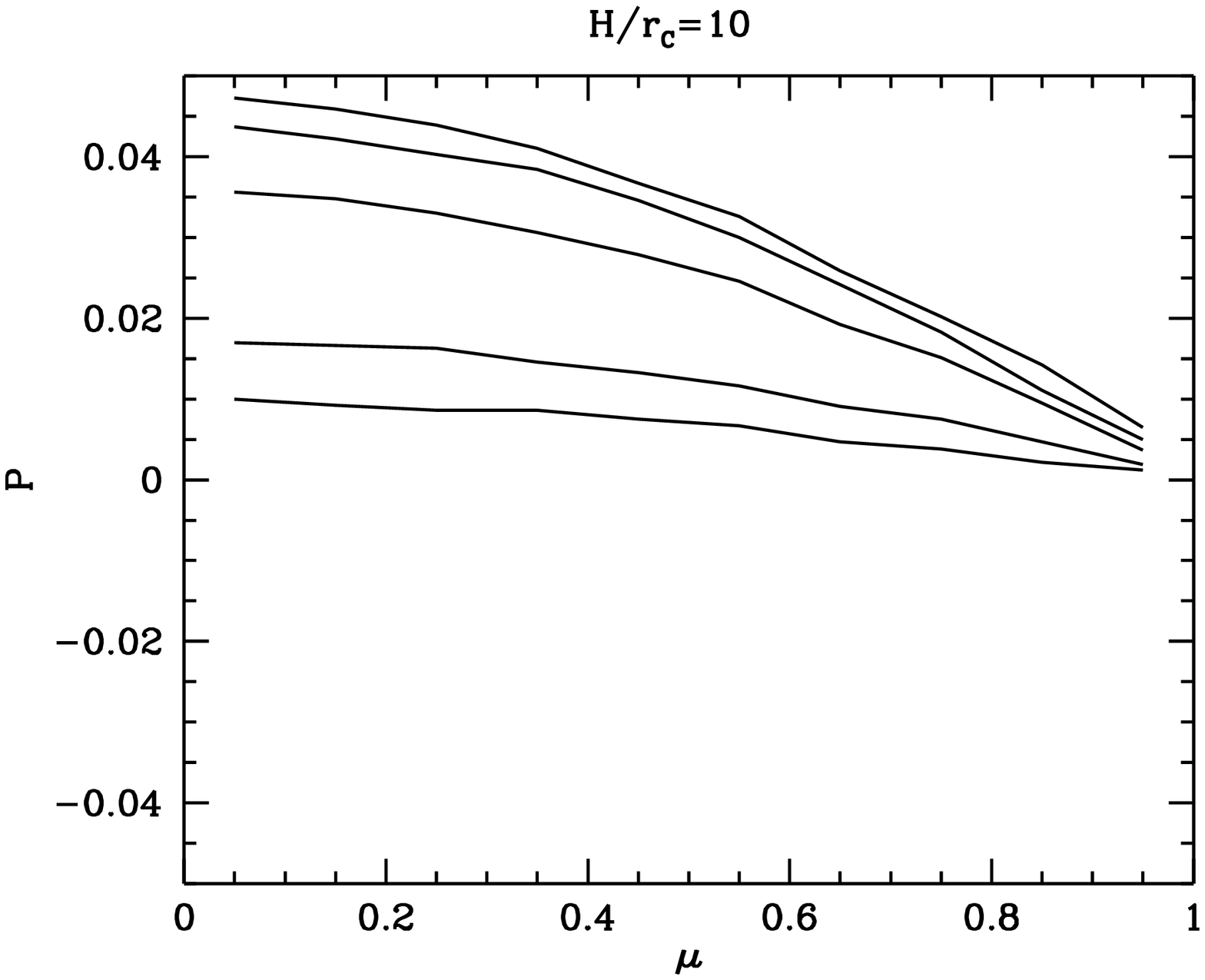}
}
\caption{The same as in the previous figure, for other four values of $H/r_C$. 
}
\label{polittico_2}
\end{figure*}

\begin{figure*}[t]
\hbox{
\includegraphics*[width=8.0cm,height=8.0cm]{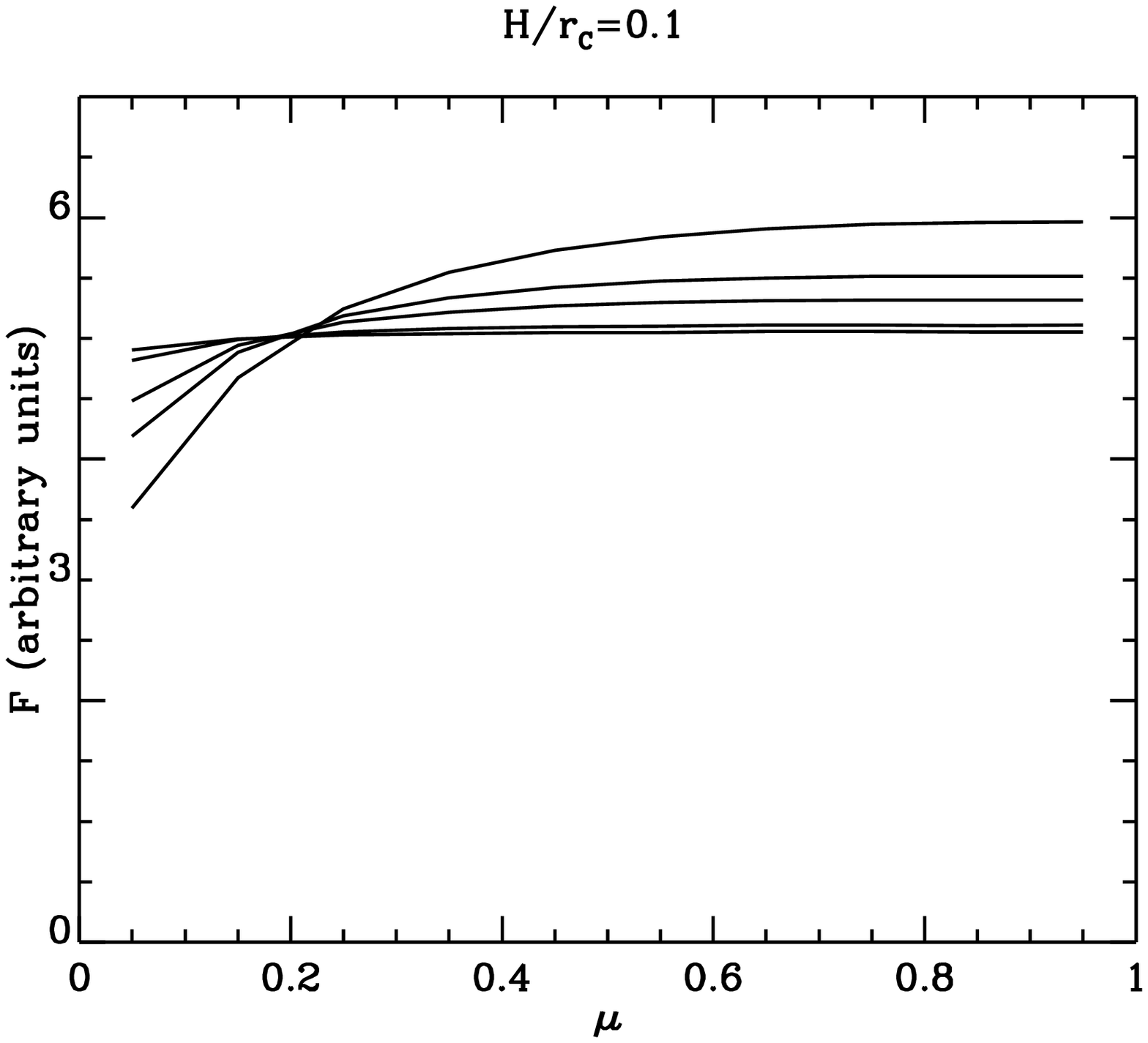}
\hspace{0.5cm}
\includegraphics*[width=8.0cm,height=8.0cm]{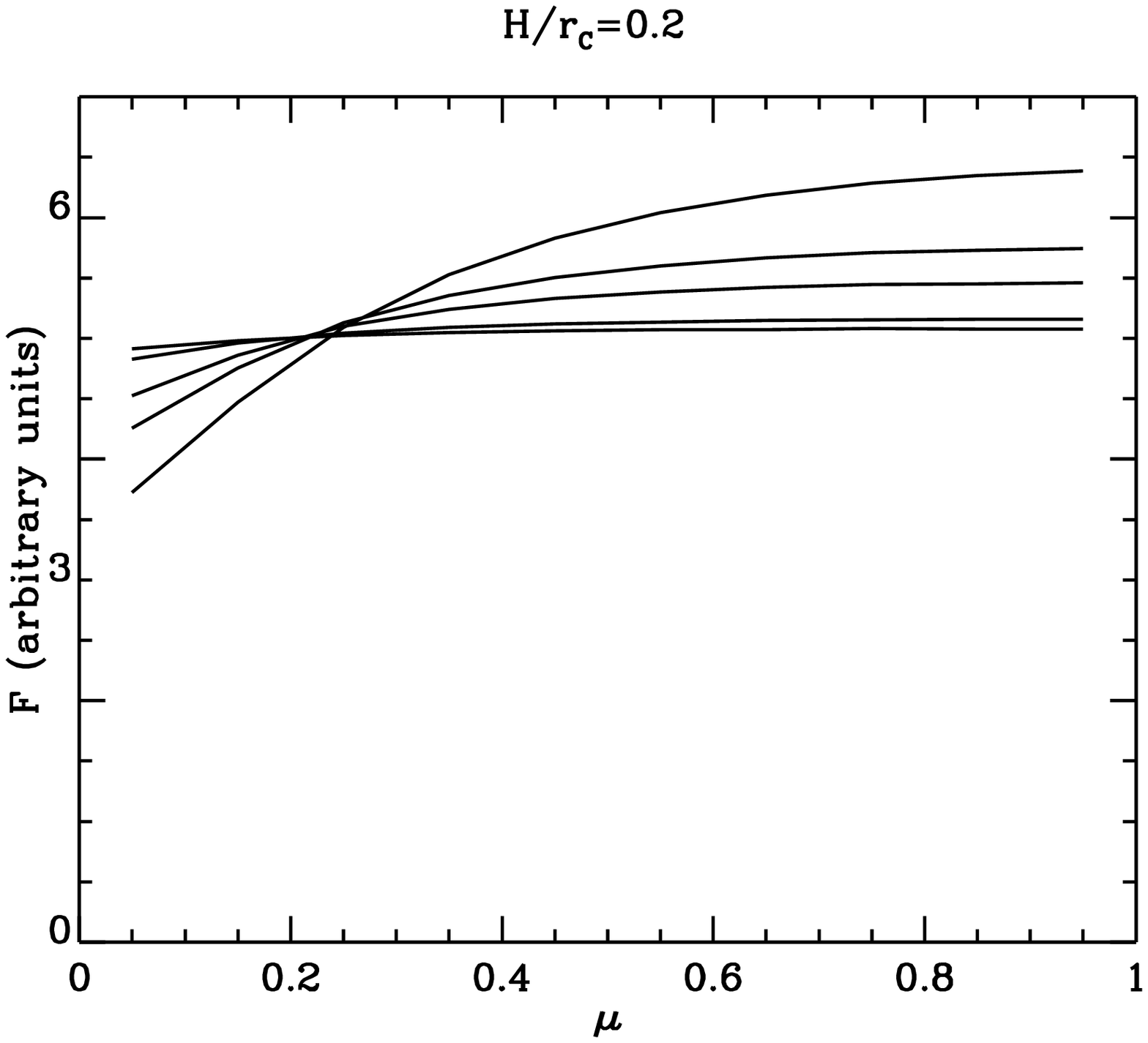}
}
\hbox{
\includegraphics*[width=8.0cm,height=8.0cm]{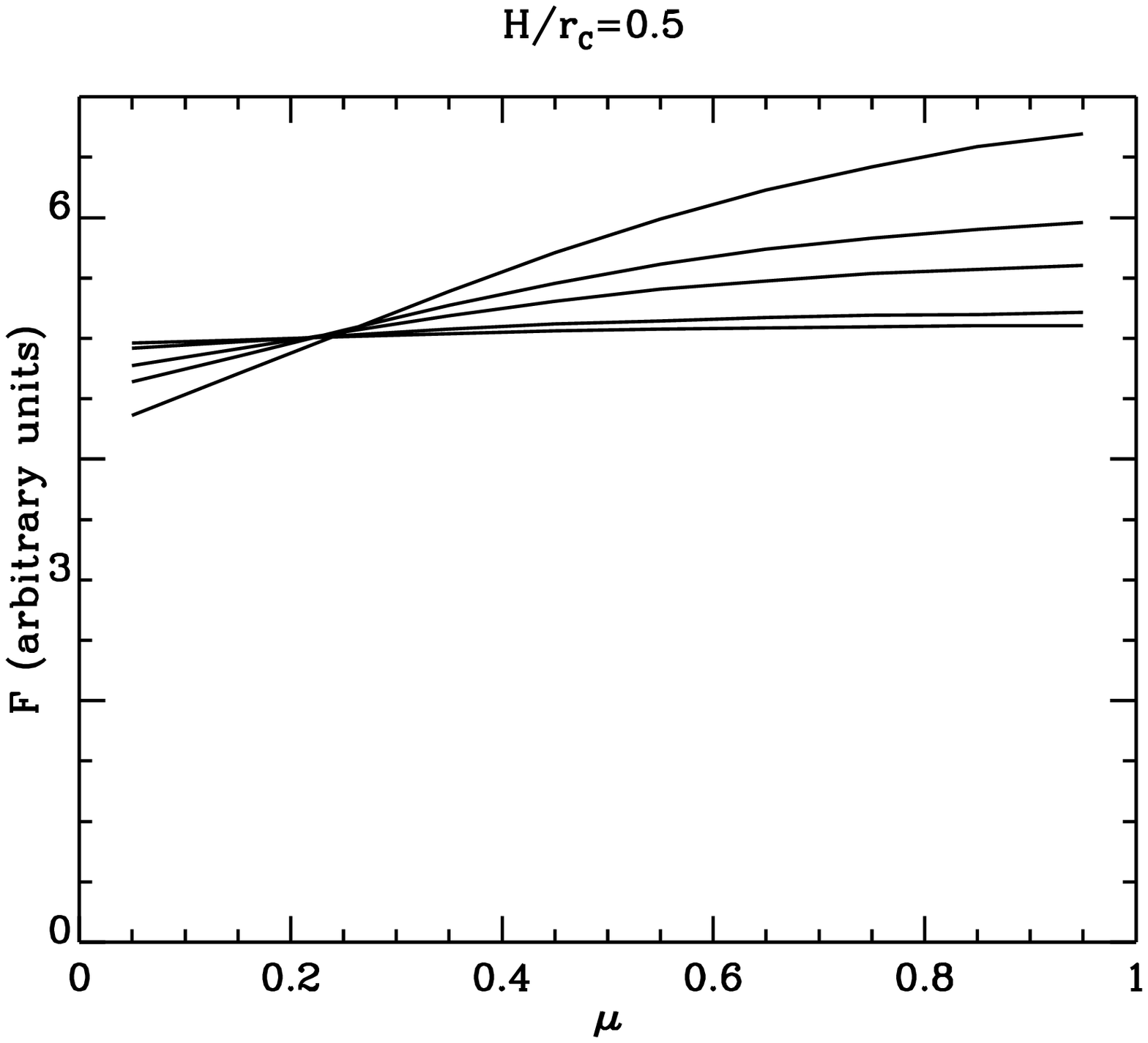}
\hspace{0.5cm}
\includegraphics*[width=8.0cm,height=8.0cm]{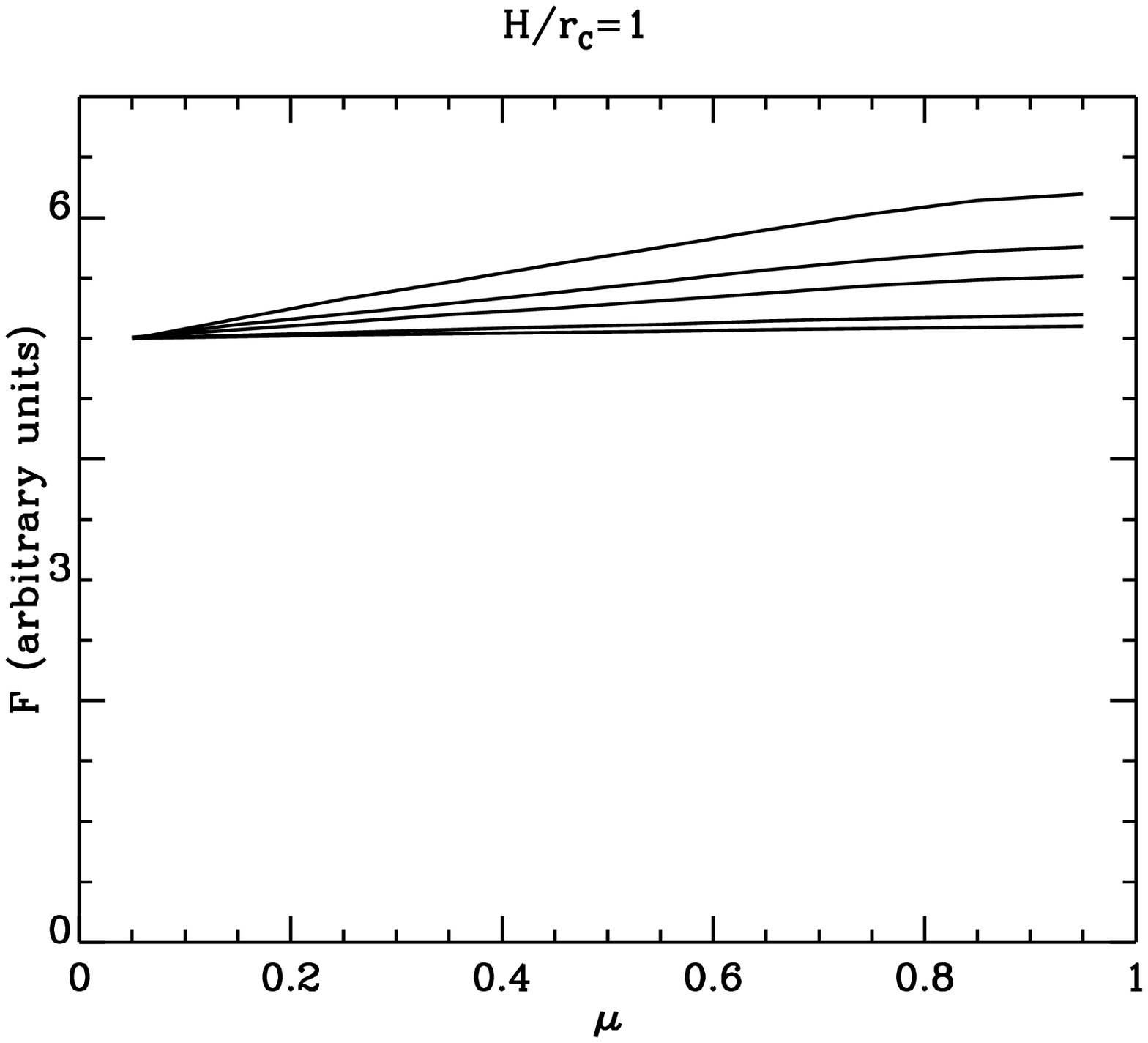}
}
\caption{The flux (in arbitrary units)  as a function of $\mu=cos\theta$ for four values
of $H/r_C$. In each panel, five values of $\tau_T$ are shown: 0.05, 0.1, 0.3, 0.5 and 1
(in increasing order of anisotropy). 
}
\label{polittico_flux_1}
\end{figure*}

\begin{figure*}[t]
\hbox{
\includegraphics*[width=8.0cm,height=8.0cm]{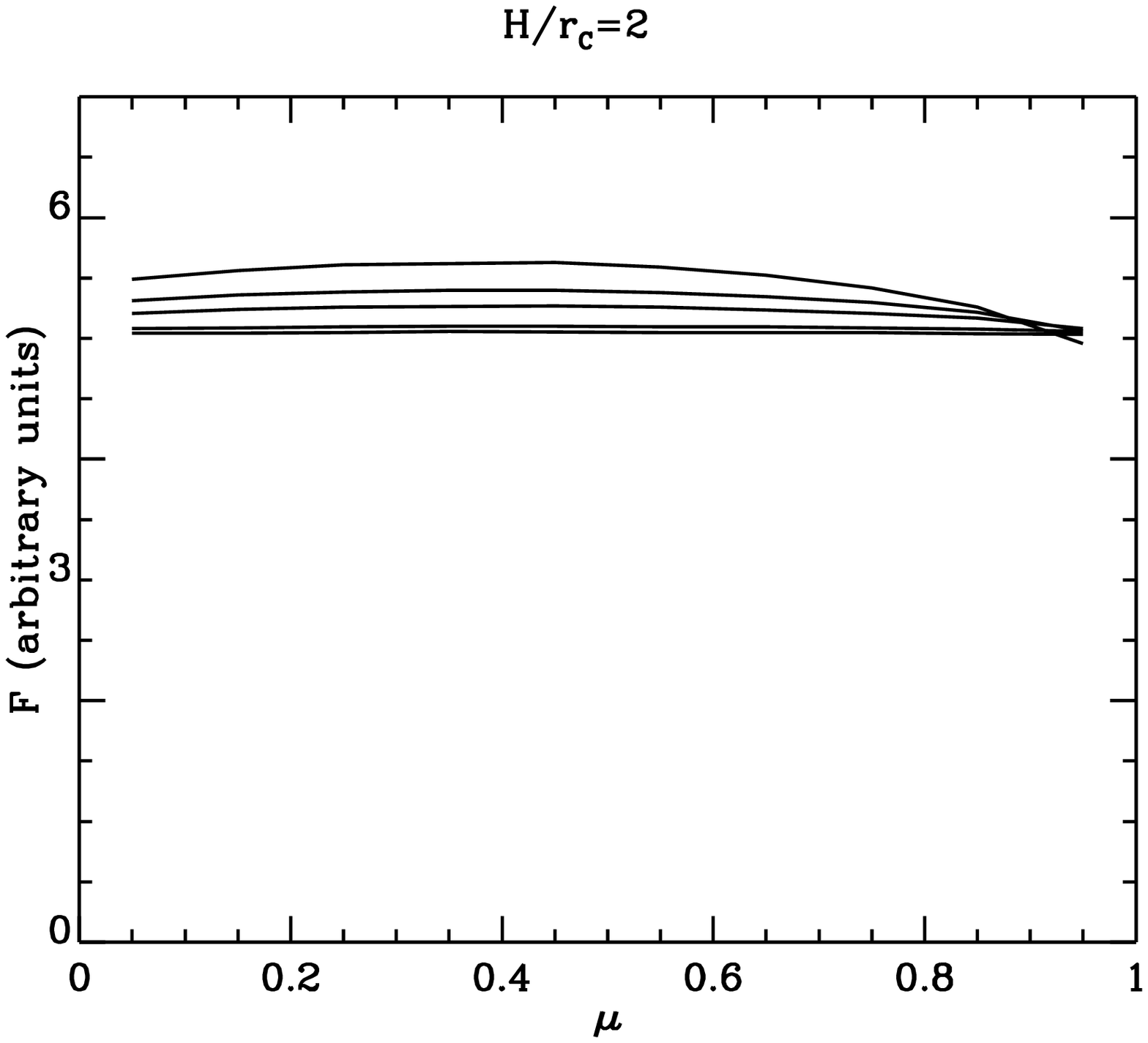}
\hspace{0.5cm}
\includegraphics*[width=8.0cm,height=8.0cm]{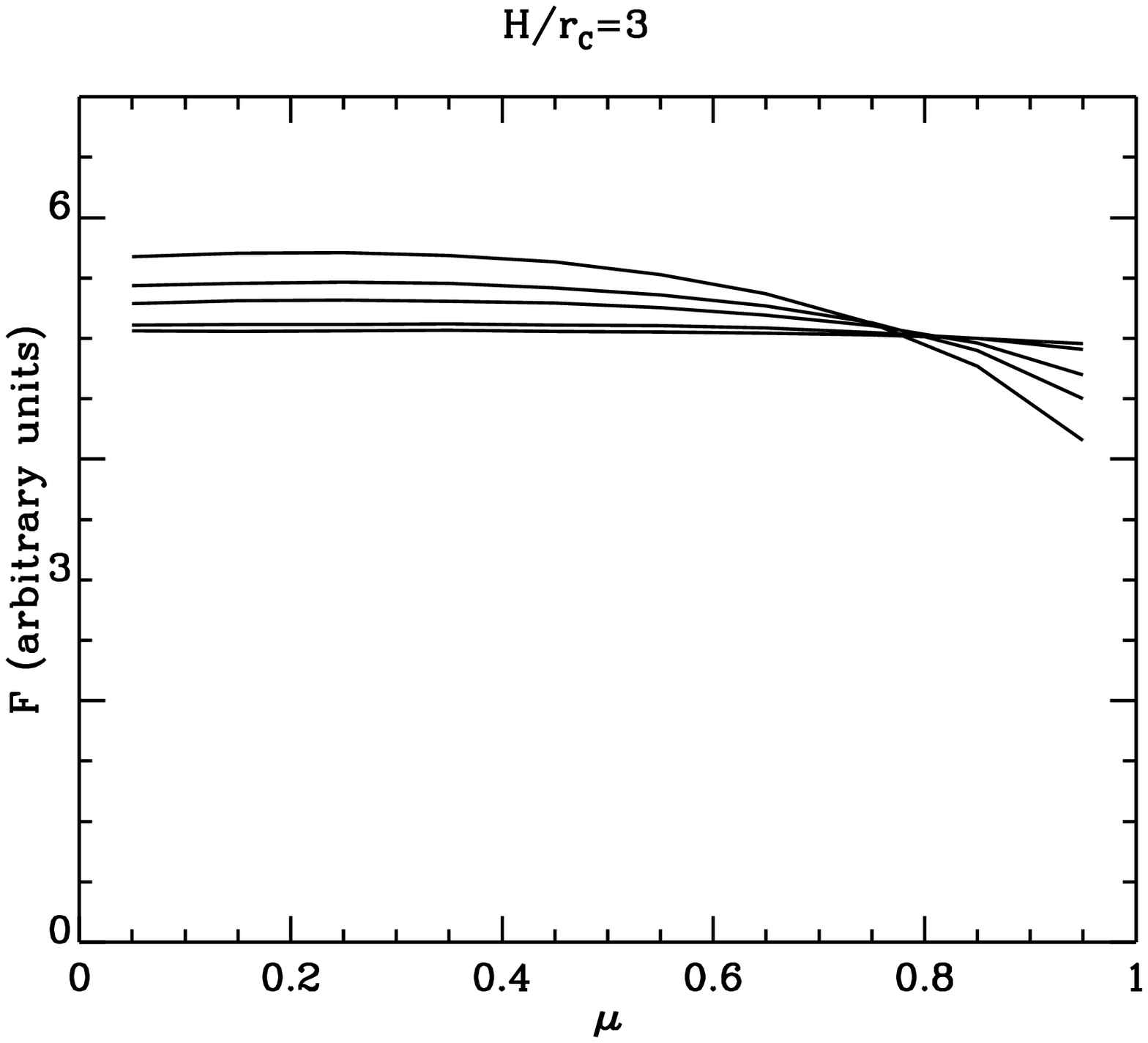}
}
\hbox{
\includegraphics*[width=8.0cm,height=8.0cm]{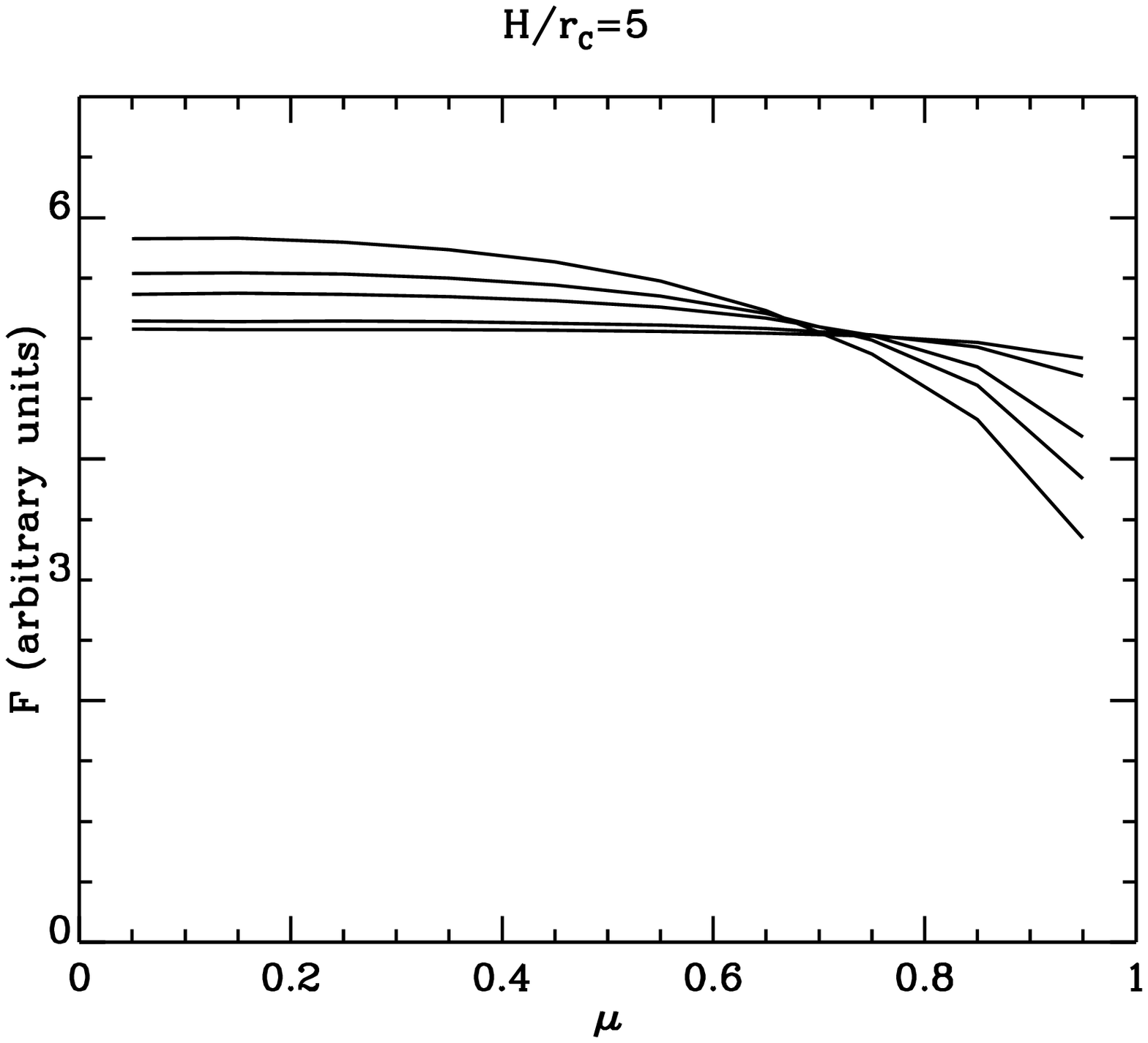}
\hspace{0.5cm}
\includegraphics*[width=8.0cm,height=8.0cm]{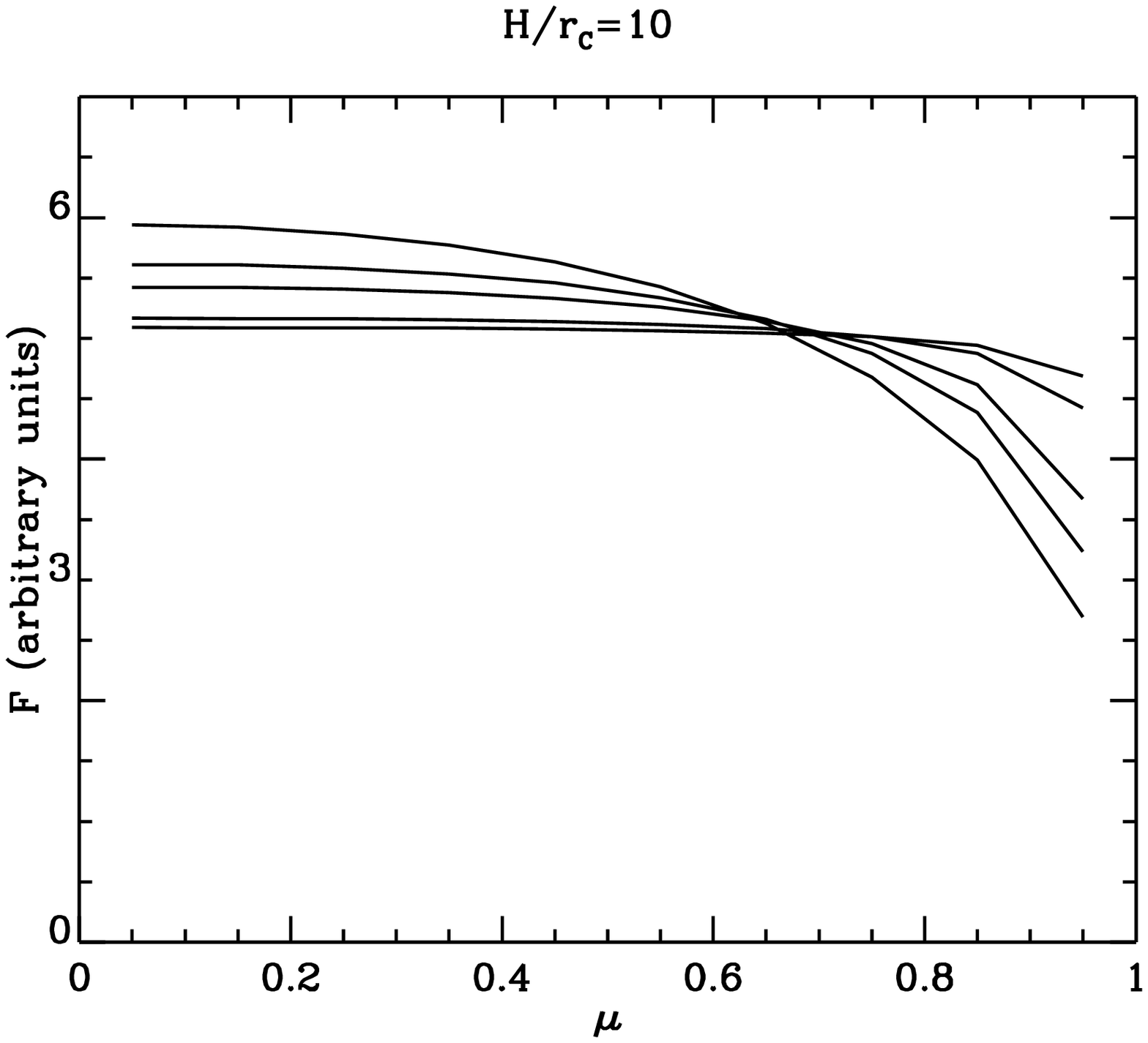}
}
\caption{The same as in the previous figure, for four other values 
of $H/r_C$. 
}
\label{polittico_flux_2}
\end{figure*}

\medskip

\section{Calculations}

The Monte Carlo technique adopted in our code is described in Matt (1993) and Matt et al. (1989;
1996). A full Compton scattering matrix is adopted (McMaster 1961). However, 
in the energy range under
consideration ($\simlt$20 keV), the scattering matrix basically 
reduces to the Rayleigh one. We assumed that primary photons (i.e. photons before
scattering) are unpolarized. Bremsstrahlung photons have a  
polarization, i.e. electric, vector perpendicular to the interaction plane,
and the random velocity field ensures that the radiation has a null net polarization. In practice,
in the code each photon is assigned a random polarization
vector. After emitted, the photon path
is followed until it leaves the accretion column. We assumed total ionization of the matter, i.e.
no photoelectric absorption.

We assumed, for simplicity, a cylindrical geometry, constant density (and therefore
emissivity) and zero temperature of the electrons. While the first
assumption, as any geometrical assumption, is critical for polarization calculations,
the second assumption is not very relevant, the important parameter being the total
scattering optical depth. The third assumption is apparently very strong -- typical
maximum temperatures of magnetic CVs are 10-20 keV for polars, and up 
to 30-40 keV for Intermediate Polars. However, especially for polars and for unpolarized
primary emission, the zero temperature approximation is still reasonably good as far as
polarization properties (see e.g. Poutanen \& Vilhu 1993) and intensity distributions are
concerned.  

We calculated the polarization properties as a function of two parameters: the radial
Thomson optical depth ($\tau_T$), and the $H/r_C$ ratio, where $r_C$ is the radius
and $H$ the height of the cylinder. Simple modeling of the accretion column gives, for 
these two parameters, the following equations (e.g. Frank et al. 1993):

\begin{equation}
H/r_C \simeq 4.5 \dot{M}^{-1}_{16} f^{1 \over 2}_{-2} M^{3 \over 2}_{wd,1} R^{-{1 \over 2}}_{wd,9}
\label{ratio}
\end{equation}

\begin{equation}
\tau_T \simeq 0.08 \dot{M}_{16} f^{-1}_{-2} M^{-{1 \over 2}}_{wd,1} R^{-{1 \over 2}}_{wd,9}
\label{tau}
\end{equation}

\noindent
where $\dot{M}_{16}$ is the accretion rate in units of 10$^{16}$ g s$^{-1}$, $f_{-2}$
is the fractional area of the column on the White Dwarf surface
in units of 10$^{-2}$ ($f$=$r_c^2$/4$R_{wd}^2$), 
$M_{wd,1}$ the mass of the White Dwarf in units of the solar mass, and $R_{wd,9}$ the  
radius of the White Dwarf in units of 10$^9$ cm. It is worth noticing that $\tau_T$
increases with the accretion rate, and therefore we expect a larger polarization degree 
when the sources are in their high states. It must also be recalled that eq.~\ref{ratio}
is valid for bremsstrahlung cooling only. If cyclotron cooling is not negligible, eq.~\ref{ratio}
gives only an upper limit to $H/r_C$ (Frank et al. 1993)..

To decide the range of parameters to be explored, we assumed $R_{wd,9}$=1 and $M_{wd,1}$=1,
and took the results on the accretion rate and column radius given in Cropper et al. (1998).
These parameters were obtained by fitting GINGA spectra with a more complex model, and 
therefore our procedure is not fully self--consistent, but certainly good enough for an
order--of--magnitude estimate. 

$\tau_T$ results to be typically in the range 0.05-1. This is therefore the range we will
explore (in any case, for $\tau_T\simlt$0.05 the polarization degree is negligible).

$H/r_C$ is typically in the range 1--10 (apart for sources in low states, when it can became
much larger but the optical depth is very low). Given the fact that $H/r_C$ is an upper limit,
we extend the explored range down to 0.1. It must be noted that for $H/r_C$ larger than 10
and smaller than 0.1, the polarization properties will not change much any further
because the geometry `saturates', apart from
very small (in the former case) or large (in the latter case) inclination angles when border
effects are still present.

Finally, it is important to note that the two parameters are not independent of each other, 
but $H/r_C$ increases with $f$
and decreases with $\dot{M}$, while $\tau_T$ has the opposite behaviour. We therefore
expect large radial optical depths for flattened geometrical configurations. For the sake
of completeness, however, we will explore the full parameter space.

1 billion input photons are used for each simulation. The emerging photons are stored
in angular bins, with $\Delta\mu$=0.1 ($\mu=cos\theta$, where $\theta$ is the 
inclination angle measured from the cylinder axis, i.e. the angle bewteen the magnetic axis
and the line--of--sight).

\section{Results}

The hard (i.e. above $\sim$1 keV) X--ray emission of magnetic CVs consists not only of
plasma (bremsstrahlung) emission but also of a reflection component, arising from the
illumination of the WD surface (e.g. van Teeseling et al 1994; Beardmore et al 1995; 
Matt 1999 and references therein).

Let us first present the polarization properties of the accretion column alone, and afterwards
we will discuss the contribution of the reflection component.

\subsection{Emission from the accretion column}

As long as the scattering matrix does not differ too much from the Rayleigh one, the
polarization properties of the X--ray radiation from the accretion column are
energy--independent. For obvious symmetry reasons, the polarization is expected
to be either parallel or perpendicular to the projection of the cylinder axis
onto the plane of the sky. In the former case, as
customary when dealing with axisymmetric geometries, we
will conventionally assume that the polarization degree, $P$,  is negative, while
positive of course in the latter case. 

In Figs.~\ref{polittico_1} and \ref{polittico_2}, 
the polarization degree as a function of $\mu$ for different
values of $\tau_T$ and $H/r_C$, is shown. For $H/r_C\simlt$1, the polarization is negative,
as expected for a disc--like geometry and $\tau_T$ less than a few 
(e.g. Sunyaev \& Titarchuk 1985). Of course, $P$ increases (in absolute value) with $\tau_T$,
as more and more photons are scattered (and therefore polarized) before escaping. $P$ 
at first decreases
with increasing $H/r_C$,  becoming very small when this parameter is equal to 1, not 
suprisingly as in this case the geometry is rather symmetric. 
Further increasing $H/r_C$, the polarization
degree  (now positive) starts to increase again.

In Figs.~\ref{polittico_flux_1} and \ref{polittico_flux_2}, 
the flux (in arbitrary units) as a function of $\mu$ is instead 
shown. The flux anisotropy of course increases with increasing $\tau_C$. 

It is worth recalling that the X--ray polarization is expected to be phase--dependent,
because of the dependence on the spin 
phase of $\theta$, which is basically the angle between the line
of sight and the magnetic dipole axis. It must be also remarked that $\theta$ can become larger
than 90$^{\circ}$, in which case the accretion column is partly occulted by the WD surface.
The degree of polarization in this case depends of course on the level of occultation, but
it is expected to be larger than in absence of occultation, because the geometry becomes
less symmetric. 

\subsection{The reflection component}

\begin{figure}
\epsfig{file=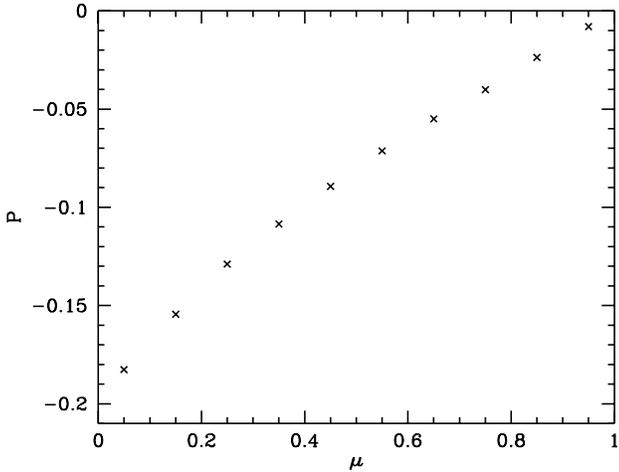,height=9.cm,width=9.cm}
\caption{The polarization degree of the reflected component, as a function of $\mu$.}
\label{polrefl}
\end{figure}

The reflection component is polarized  parallel to the projected cylinder axis (Matt et.
1989; Matt 1993), i.e. it is negative. The degree of polarization as a function of $\theta$ is
shown in Fig.~\ref{polrefl}, again obtained by means of Monte Carlo simulations (Matt et al. 1989;
photons are integrated over the 1-20 keV energy range). The net
polarization is then given by:

\begin{equation}
P(E,\mu) = {P_{pr}(\mu) + r(E,\mu) P_{ref}(E,\mu) \over 1 + r(E,\mu) } $$
\end{equation}

\noindent
where $P_{pr}$ and $P_{ref}$ are the polarization degrees of the primary and reflected
emission, respectively (taken with their signs), and $r(E,\mu)$ is the ratio between
the primary and reflected fluxes (see e.g. Matt 1999). 

\subsection{The expected X--ray polarization of AM Herculis}. 

As an application of our calculations, let us discuss
the expected phase--dependent X--ray polarization of the archetypal polar, AM Herculis. 
Assuming the system parameters given by Cropper (1988), i.e. $\beta$=61$^{\circ}$ 
and $i$=30$^{\circ}$ (where $\beta$ is the angle between the magnetic and spin axes, 
and $i$ the angle between the spin axis and the line--of--sight),
$\theta$ never exceeds significantly
90$^{\circ}$, so our calculations can be safely applied to the whole spin period. 

\begin{figure}
\epsfig{file=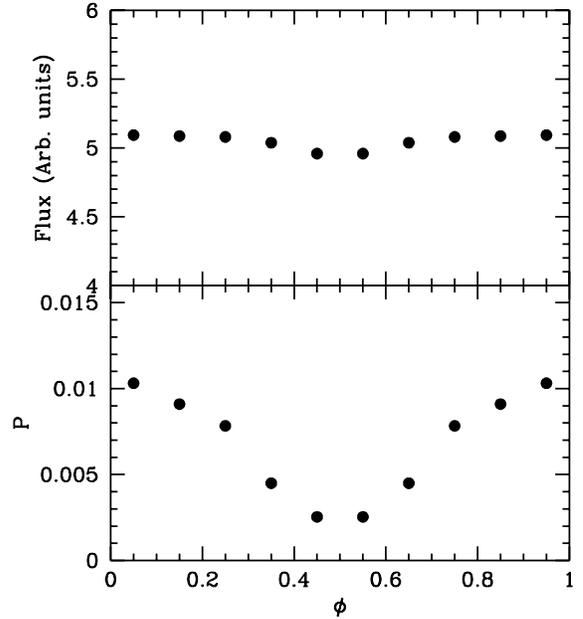,height=10.cm}
\caption{The intensity (upper panel) and degree of polarization (lower panel) expected for
AM~Herculis, calculated without the reflection component, as a function
of the spin phase. $H/r_C$=10 and $\tau_T$=0.05.}
\label{amher00510}
\end{figure}

\begin{figure}
\epsfig{file=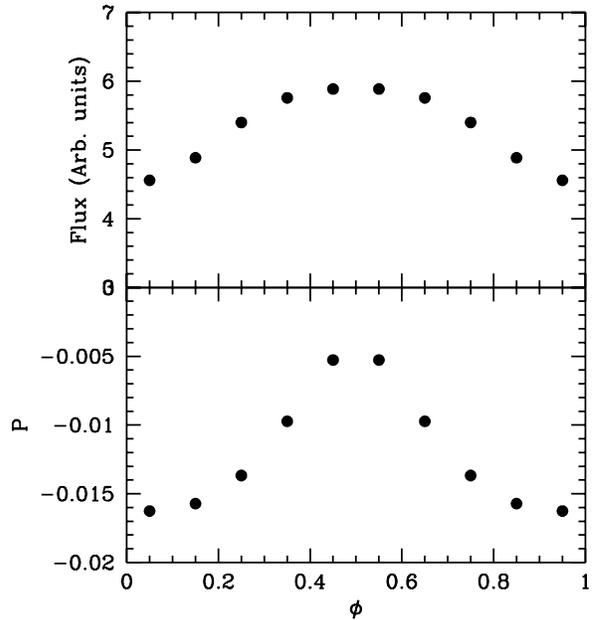,height=10.cm}
\caption{The intensity (upper panel) and degree of polarization (lower panel) expected for
AM~Herculis, calculated without the reflection component, as a function
of the spin phase. $H/r_C$=0.5 and $\tau_T$=0.5.}
\label{amher0505}
\end{figure}

The intensity and degree of polarization of the accretion column on the main magnetic pole, 
without the reflection component (and 
therefore valid for energies less than a few keV), as a function
of the spin phase, $\phi$, are shown in Figs.~\ref{amher00510} and ~\ref{amher0505}. 
In Fig.~\ref{amher00510}, $H/r_C$=10 and $\tau_T$=0.05
have been assumed, according to the value derived from the Cropper et al. (1998) estimates
from ASCA data.
However, during the ASCA observation the source was in an intermediate X--ray state; larger
optical depths, and lower $H/r_C$, are expected at higher luminosities. Moreover, it must be 
recalled again that the  $H/r_C$ values estimated from eq.~\ref{ratio} are upper limit, as 
cyclotron cooling is not taken into account. We therefore calculated the phase--dependent
polarization properties also for the (more favourable) case of $H/r_C$=0.5 and $\tau_T$=0.5
(Fig.~\ref{amher0505}). 

In the first case (Fig.~\ref{amher00510}), the polarization degree is low (less than 1\%)
because of the low optical depth. The flux is basically phase--independent, differently
from what found in intermediate states by BeppoSAX and ASCA (Ishida et al. 1997, Matt et
al. 2000). Unless the real optical depth is larger than derived from eq.~2, this means that
either the assumed geometrical and physical parameters of the accreting column 
are oversimplified (it must be recalled that we have assumed a single temperature and density
along the column), or that there is substantial phase--dependent absorption. Studying deviations of
the measured polarization from the expected one may help distinguishing between the different
hypotheses.

The phase--dependent flux shown in Fig.~\ref{amher0505} is instead 
roughly similar to that observed by BeppoSAX during a high state of the source (Matt et al. 2000);
the observed amplitude of the modulation is however larger then predicted, again suggesting a more
complex gemetrical and/or physical situation than assumed in our calculations.
The polarization degree is expected to be as large as 1.5\%, increasing
when the flux decreases.

\begin{figure}
\epsfig{file=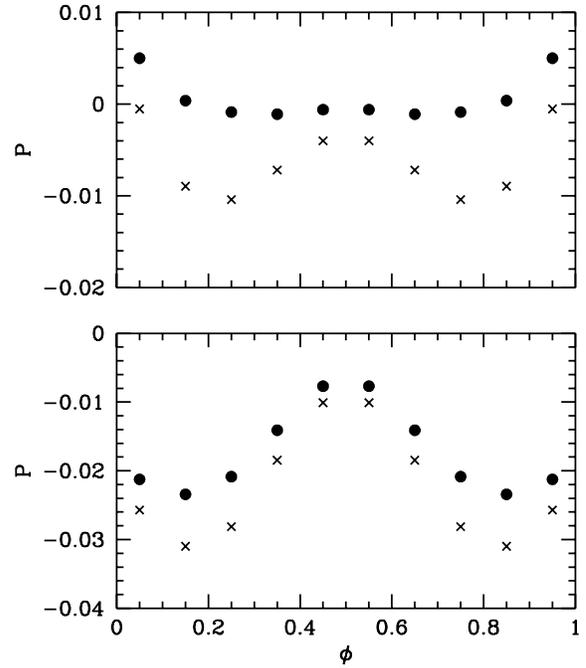,height=10.cm}
\caption{The degree of polarization, including the reflection component, expected for
AM~Herculis
as a function of the spin phase, for two energy bins (5--10 keV, filled circles,
and 10--15 keV, crosses). Upper panel: 
$H/r_C$=10 and $\tau_T$=0.05; lower panel: $H/r_C$=0.5 and $\tau_T$=0.5.}
\label{amher_ref}
\end{figure}

The phase--dependent degree of polarization, including the reflection
component as estimated by Matt et al. 
(2000) during a high state of the source, is shown in Fig.~\ref{amher_ref} 
 for two energy bins (5--10 keV and 10--15 keV); $R_C/H$=10 and $\tau_T$=0.05 (upper panel)
and $R_C/H$=0.5 and $\tau_T$=0.5 (lower panel).

\section{Summary}

We have calculated the polarization properties of the accretion column in magnetic CVs.
Polarization arises because a fraction of the thermal radiation can be Thomson scattered
before escaping the column. The polarization degree can be as high as $\sim$4\%, and depends
on the angle between the magnetic field and the line--of--sight (which of course varies
with the spin phase), which is often well known (e.g. Cropper 1988). It depends also on
the Thomson optical depth, $\tau_T$, and on the ratio between the radius and height
of the accretion column. If one of these two parameters can be independently estimated
(e.g. $\tau_T$ from the iron line broadening, e.g. Hellier et al. 1998), than the other
can in principle be deduced by polarization measurements. In particular, as the polarization
degree is negative (positive) for $H/r_C$ less (greater) than one, while the polarization
of the reflection component (whose importance increases with energy) is always negative,
an increase (decrease) of the polarization degree with energy is expected in the former
(latter) case.
 
In this paper we have discussed the continuum emission. However, iron lines
provide a significant fraction of the total X--ray flux. Recombination lines emitted in the
accretion column may suffer not only Compton scattering but also resonant scattering,
so the effective optical depth in the line is larger than in the continuum. Line photons
are therefore likely to be more polarized than continuum photons. 
On the contrary, the fluorescent neutral line 
emitted by the White Dwarf surface is unpolarized, at least in the line core.

Degrees of polarization of the order of one percent 
are within the detection capabilities of the new
generation X--ray polarimeters based on the photoelectric effect 
(Costa et al. 2001) when coupled with large enough X--ray telescopes
(Costa et al. 2003), at least for the brightest magnetic CVs
in high state (when their flux can be as high as several millicrabs;
for instance the 2--10 keV flux of AM~Herculis in high state is about 10$^{-10}$ 
erg cm$^{-2}$ s$^{-1}$, Matt et al. 2000).  Moreover, the spectral resolution
should be good enough to search for different polarization degrees in the emission 
iron lines. Bright magnetic CVs should therefore be 
added to the traditional lists of targets for future polarimetric missions.

\section*{Acknowledgements}
I thank Domitilla de Martino and Koji Mukai for very useful comments
and suggestions. Financial support from ASI is acknowledged.

\end{document}